\documentclass[conference]{IEEEtran}
\IEEEoverridecommandlockouts
\usepackage{cite}
\usepackage{amsmath,amssymb,amsfonts}
\usepackage{algorithmic}
\usepackage{graphicx}
\usepackage{textcomp}
\usepackage{xcolor}
\usepackage{soul}
\usepackage{enumitem}

\usepackage{fancyhdr}
\begin{document}
\bstctlcite{IEEEexample:BSTcontrol}

\title{i2Slicer: Enabling Flexible and Automated Orchestration of 5G SA End-to-End Network Slices\\
}
\author{
	\IEEEauthorblockN{
        Miguel Catalan-Cid,
        Adriana Fernández-Fernández,
        Daniel Camps-Mur,
        Shuaib Siddiqui
        \IEEEauthorblockA{
            \textit{i2CAT Foundation, Barcelona, Spain}\\
            email: \{miguel.catalan, adriana.fernandez, daniel.camps, shuaib.siddiqui\}@i2cat.net
        }
    }
}

\maketitle
\thispagestyle{fancy} 

\begin{abstract}
5G network slicing implies a step forward in customizing radio access and core networks by allowing the creation of logical networks adapted to service requirements. In addition, softwarisation has fueled the emergence of 5G solutions which do not require specialized hardware platforms. Therefore, a key requirement to drive the adoption of 5G slicing by verticals is to simplify its management through automated orchestration. In this paper, we present i2Slicer, a flexible solution to orchestrate the deployment of 5G standalone end-to-end network slices with multi-tenancy and multi-service capabilities. The implementation and evaluation of i2Slicer using state-of-the-art 5G software and hardware demonstrate that it offers a practical and efficient lifecycle management of network slices. 
\end{abstract}



\section{Introduction}
One of the main features of 5G networks is the support of customized radio access, core and transport networks adapted to the requirements of vertical services, which demand a flexible and cloud-native network architecture. In the 5G Core, this was achieved by adopting the Service Based Architecture (SBA) and Network Function Virtualisation (NFV) paradigms \cite{3gpp.23.501}. SBA and NFV allow the flexible deployment and interconnection of virtualized Network Functions (NFs) like the Access and Mobility Management Function (AMF), the Session Management Function (SMF) or the User Plane Function (UPF), further facilitating functionalities already provided by 4G, such as Multi-Operator Core Networks (MOCN) or Control User Plane Separation (CUPS). These functionalities are still relevant in 5G networks to enable multi-tenancy, network infrastructure sharing, or edge computing. However, the introduction of the slicing paradigm implied a step forward in customization and flexibility of 5G networks. 

Network slicing has consolidated as a key enabling technology in the context of 5G and beyond networks, addressing the need for network customization and service differentiation. Network slicing enables the creation of virtual, logically isolated networks, tailored to meet specific service requirements and cater to diverse use cases. End-to-end (E2E) network slices span over radio, edge and core resources on top of a common network infrastructure, which are individually manageable and interconnected to form an isolated network environment~\cite{PAPAGEORGIOU2020232}.

In recent years, softwarisation has fueled the emergence and availability of 5G RAN and Core solutions, which can run on top of general-purpose hardware and commercial off-the-shelf (COTS) equipment. In such an open ecosystem, Software Defined Networking (SDN) and orchestration frameworks are key to empowering 5G adoption by service verticals. 

To this end, in this paper, we present and evaluate i2Slicer, a practical and flexible solution to orchestrate the deployment of 5G Standalone (SA) E2E network slices using cutting-edge 5G software and hardware. i2Slicer offers multi-tenancy and multi-service capabilities by implementing MOCN, RAN sharing and 5G slicing features, which are relevant for 5G SA scenarios like private 5G or pop-up/temporary networks. The evaluation of i2Slicer demonstrates the efficiency of the developed Lifecycle Management (LCM) operations. 

\section{Background and Related Work}
5G network slicing makes use of slice identifiers to create logical networks adapted to the use case needs, which may share a common physical infrastructure. 3GPP identifies three management functions related to network slicing management, namely Communication Service Management Function (CSMF), Network Slice Management Function (NSMF) and Network Slice Subnet Management Function (NSSMF)~\cite{3gpp.28.801}. These functions provide the necessary capabilities to configure, allocate resources, and ensure the efficient operation of network slices based on specific service requirements and network conditions. A Network Slice instance (NSI) includes different Network Slice Subnets Instances (NSSI), mainly the radio access, the core and the transport networks. 

These logical networks or subnets are treated as isolated virtual instances of the 5G network, enabling network customization to meet specific service level requirements~\cite{surveyslicing}. At the core side, this allows for instance to dedicate specific NFs to selected slices and control their associated resources. At the RAN level, slicing can be considered in the radio resource management, for instance by allocating a specific amount of physical resource blocks. RAN disaggregation will allow further advance in the RAN customization, for instance by dedicating Central Units (CUs) to specific slices, as is considered in the slicing architecture adopted by the O-RAN Alliance \cite{ORAN-slicing}. At the transport network domain, SDN solutions allow to incorporate network programmability to apply, for instance, prioritization and isolation policies. 


Network slicing is closely intertwined with network softwarization, particularly NFV, to facilitate the rapid and on-demand creation of network slices to cater to diverse requirements. NFV plays a pivotal role in enabling swift and dynamic provisioning and configuration of the network services associated with the slices. 
In this regard, the Management and Orchestration (MANO) framework provides the necessary functions to manage the lifecycle of network slices~\cite{3gpp.28.801}.

In this context, several recent studies have been dedicated to the analysis and practical validation of network slice deployment. The work in~\cite{9851288} presents an experimental testbed for the management and orchestration of network slices in the core network. In each slice, an entire and independent instance of the mobile core is deployed, representing a scenario involving multiple operators, but lacking the inclusion of the RAN segment in order to achieve E2E network slices. A
similar approach for the deployment of the core network within network slices, which we refer to as \textit{monolithic} in this paper, has been proposed in our previous work, enabling multi-tenancy through MOCN~\cite{s21238103}. 

The decomposition of the core network into functional entities within the SBA introduces a more refined approach to function sharing among multiple network slices, allowing for increased deployment granularity and flexibility. 
Building upon this approach, the work presented in~\cite{9912617} demonstrates a proof of concept implementation for orchestrating network slices encompassing the transport and core domains. In the core domain, the initial slice deploys core NFs that are shared among subsequent slices. Additionally, specific NFs are deployed on a per-slice basis, ensuring customized functionality for individual network slices. In this paper, we refer to this deployment mode as \textit{disaggregated}.

Overall, it is worth noting that surveyed literature lacks the inclusion of the RAN segment in the composition of E2E network slices. Additionally, the use of RAN emulators instead of real 5G hardware is a common shortage in current research efforts. 
Complementing the aforementioned works, this paper presents a single management system that can support both multi-tenancy and multi-service within each tenant (through the disaggregated deployment mode) and provides the practical implementation of the system. The developed prototype integrates cutting-edge 5G RAN and Core solutions to create E2E network slices in the 5G SA environment.

\section{i2Slicer Design} \label{design}

\begin{figure}[ht]
\centerline{\includegraphics[width=0.5\textwidth]{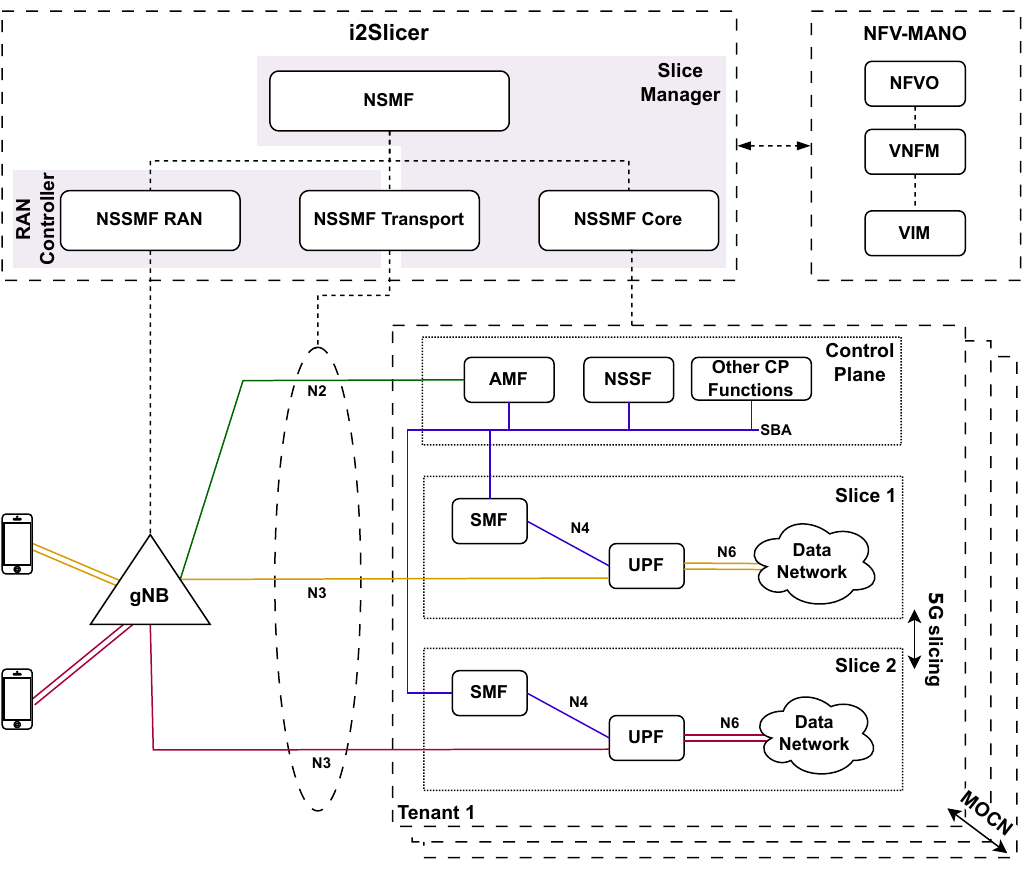}}
\caption{i2Slicer network slicing architecture}
\label{fig1}
\end{figure}



The main goal of i2Slicer is to offer multi-tenancy and multi-service capabilities. While multi-tenancy can be achieved through MOCN and RAN sharing even in the case of a monolithic deployment, a proper support of multi-service features requires of a disaggregated deployment exploiting 5G slicing to allow a dynamic and efficient management of network resources according to service status and requirements. 

To this end, as shown in Figure~\ref{fig1}, the designed disaggregated slicing approach consists of a common control plane with shared NFs for all the slices from a common operator or tenant (i.e., using a common Public Land Mobile Network ID (PLMNID)), and of an isolated user plane with dedicated SMF and UPF instances per slice. Although strictly belonging to the control plane, a dedicated SMF allows each slice to implement differentiated session management or UPF selection policies. In addition, this allocation approach isolates the user plane operation of different slices in case of SMF malfunctions or service interruption. Regarding the control plane, the figure depicts the AMF and the Network Slice Selection Function (NSSF) since these NFs are also involved in the definition and selection of slices \cite{3gpp.23.501}. Finally, note that multi-tenancy is supported through MOCN by deploying additional control planes plus dedicated user plane slices. 

As depicted in Figure~\ref{fig1}, in this work we focus on the NSMF and NSSMF layers, whose main functionalities include:

\begin{itemize}
  \item NSMF: 
  handles the creation, modification, and termination of network slices based on related requirements. The NSMF derives constituent Network Slice Subnets (NSSs) and interacts with the underlying NSSMFs to allocate resources, configure NFs, and enforce policies specific to each slice. 
  \item NSSMF: manages the lifecycle of NSSIs, which are portions of the network infrastructure dedicated to specific network slices. The NSSMF is responsible for the configuration and control of the subnet resources 
  to meet the demands of the associated network slice.
\end{itemize}


As per the 3GPP specifications ~\cite{3gpp.28.533}, 
the NSSMF is capable of consuming the LCM services provided by the NFV-MANO, specifically the NFV Orchestrator (NFVO). Simultaneously, the NSSMF also serves as a producer of the management services associated with the network slice subnet. Through these interactions, the management functions within the network slicing architecture seamlessly integrate with the NFV-MANO framework, allowing for efficient control and coordination of the network slices, including associated virtualized applications and services.

For the implementation of the slice management components, we have utilized our own developed tools, namely the Slice Manager and the RAN Controller. These tools have been successfully demonstrated in several H2020 projects such as 5G-VICTORI\footnote{https://www.5g-victori-project.eu/}, where we developed and integrated the approach being presented in this paper \cite{5gvictori}. 

The Slice Manager plays a pivotal role as the NSMF, being responsible for orchestrating the overall lifecycle of E2E network slices. Moreover, it functions as the NSSMF Core, handling the deployment and configuration of the core network over the virtualized infrastructure. The Slice Manager interacts with the RAN Controller, which implements the NSSMF RAN, allowing dynamic and remote configuration of the RAN nodes through the NETCONF/O1 protocol. Additionally, both components ensure the successful configuration of networking paths within the transport domain. This involves setting up L2 networks, particularly VLANs, to guarantee the necessary isolation between the control, user and data planes of the deployed network slices (i.e., N2, N3 and N6 interfaces). In the case of the data plane, application functions may be deployed and interconnected to the specific VLAN of the slice or service.

For the LCM of network slices, i2Slicer implements the key operations as described below:

\begin{itemize}
  \item Slice creation: entails provisioning the necessary logical entities over the infrastructure resources required to support the network slice. This includes the creation of a slice-specific tenant and the establishment of the required VLAN networks, ensuring slice isolation. 
  These steps lay the foundation for the subsequent stages.
  \item Slice activation: encompasses the configurations specific to the RAN and Core slice subnets, plus the deployment of the core instances. The 5G Core deployment varies depending on the chosen deployment mode, since both monolithic and disaggregated approaches are supported by i2Slicer, as follows:
      \begin{enumerate}[label=\alph*)]
        \item Monolithic mode: A single instance with all NFs is deployed per tenant.
        \item Disaggregated mode:
            \begin{enumerate}[label=(\roman*)]
                \item For the first slice of the tenant, three instances are deployed: the shared control plane functions, the SMF and the UPF.
                \item For subsequent slices, only the two slice-specific NFs are deployed: SMF and UPF. 
        \end{enumerate}
      \end{enumerate}
  \item Slice removal: involves terminating all services and logical entities associated with the slice. This phase ensures the clean removal and release of the allocated resources.
\end{itemize}

\section{i2Slicer Evaluation}

In this section, we showcase the experimental evaluation of i2Slicer. The purpose is to compare the monolithic and disaggregated deployment modes in terms of efficiency, analysing their execution times and resource utilization. 

The developed prototype is based on two well-known solutions from the state-of-the-art, namely: Open5Gs and Amarisoft. Open5Gs\footnote{https://open5gs.org/} is an open-source implementation of the 5G Core (Release 17) which supports all required functionalities for our slicing approach such as SBA, NFV, CUPS and MOCN. It also allows a flexible configuration of the different NFs through YAML files, which in our implementation are managed by the Slice Manager. On the other hand, the Amarisoft Callbox\footnote{https://www.amarisoft.com/products/test-measurements/amari-lte-callbox/} acts as a 3GPP compliant gNB (release 16) using Software Defined Radios (SDRs) that offer limited coverage but a notable performance \cite{amarisoft}. As aforementioned, we implemented the NETCONF protocol to allow remote radio configuration and control through our RAN Controller. As for the NFV-MANO framework, we have leveraged the Open Source MANO (OSM) (release 11), which operates as the NFVO and VNFM. Additionally, Openstack (release Victoria) has been employed as the VIM.

\begin{figure}[t!]
\includegraphics[width=1.0\linewidth]{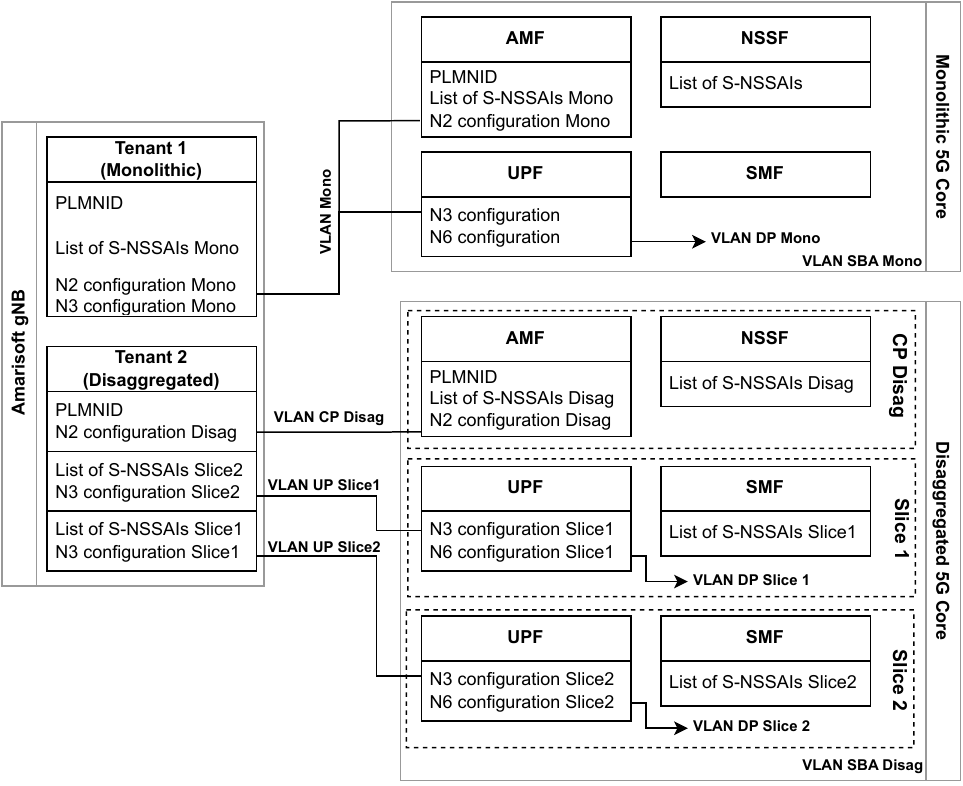}
\caption{Network slicing example: main configurations and VLANs}
\label{fig2}
\end{figure}

Figure~\ref{fig2} illustrates the main parameters configured by i2Slicer during monolithic and disaggregated slice deployments: PLMNIDs, S-NSSAIs and N2/N3/N6 configurations. Additionally, the required VLANs are created according to the deployment mode. Note that in the case of the disaggregated 5G Core, the SMFs include the list of the S-NSSAIs being served by each slice; this is used by the AMF to select the required SMF for each PDU session. For simplicity, Figure~\ref{fig2} does not include radio configurations such as the frequency or the bandwidth, which can also be set through our RAN Controller prior to the slice deployment. 

Table~\ref{tab1} outlines the minimum resources required by the considered deployment modes. The amount of RAM is based on the minimum requirement of 4 GBs for MongoDB, which is used by some NFs in the Open5Gs control plane; on the other hand, SMF and UPF instances just need 1 GB. In the case of the number of virtual CPUs, the UPF needs 2 while the SMF and CP instances can work with only 1.  

\begin{table}[htbp]
\caption{Minimum Required Resources}
\begin{center}
\begin{tabular}{|c|c|c|}
\hline
\textbf{Deployment Mode} & \textbf{\textit{\# vCPU}} & \textbf{\textit{RAM (GB)}}\\
\hline
Monolithic (per tenant/slice) & 4 & 4 \\
\hline
Disaggregated (First slice per tenant) & 4 & 6 \\
\hline
Disaggregated (Subsequent slices) & 3 & 2 \\
\hline
\end{tabular}
\label{tab1}
\end{center}
\end{table}


Considering the values presented in Table~\ref{tab1}, 
it is evident that, despite of the resources needed for the first disaggregated slice, which entails the deployment of the control plane plus the SMF and the UPF instances, the subsequent slices offer a more efficient solution than the monolithic case. This results in reductions of up to $25$\% in vCPU usage and $50$\% in RAM usage. This efficiency becomes increasingly significant as the number of slices increases, showcasing the scalability and cost-effectiveness of the disaggregated mode in managing resource allocation. 

\begin{figure}[t!]
\centering
\includegraphics[width=.78\linewidth]{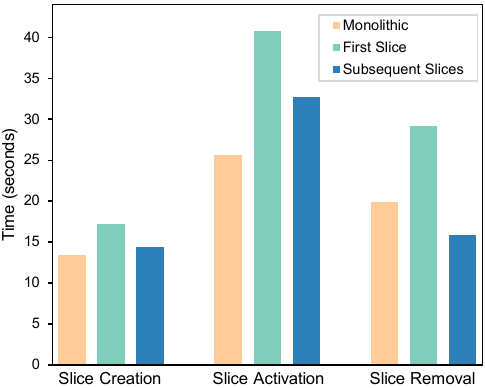}
\caption{Time spent during slice management operations}
\label{fig3}
\end{figure}

The results presented in Figure~\ref{fig3} correspond to the average execution time for each slice management operation over a total of $100$ iterations. As can be observed, the disaggregated approach exhibits longer deployment times due to the additional creation of logical resource entities and NFs, especially during the slice activation phase (the impact of RAN configuration operations was negligible). As was described in the previous section, in this case, a larger number of core network function instances need to be deployed, while the monolithic approach only requires a single instance. However, it is worth noting that the impact decreases for the case of subsequent slices since the control plane is no longer redeployed. In any case, note that all the evaluated cases achieved execution times below $45$ seconds, which are within reasonable limits for scenarios such as temporary or pop-up networks. 

These findings indicate that, albeit i2Slicer's disaggregated approach incurs slightly higher deployment times,
it is an acceptable trade-off for its main benefits: traffic and resource isolation, finer resource allocation, and higher functions placement flexibility for multi-service capabilities. 
This allows tenants to efficiently manage shared resources according to service status and requirements instead of over-provisioning monolithic deployments. 

\section{Conclusions}
This paper presented i2Slicer, a practical approach to orchestrate 5G SA E2E network slices using cutting-edge 5G RAN and Core solutions, namely Amarisoft and Open5Gs, using our own developed NSMF and NSSMF tools. i2Slicer offers a flexible multi-tenant and multi-service architecture supporting features such as 5G SA slicing, MOCN, and the separation and isolation of control, user and data planes. The evaluation demonstrated the efficiency of i2Slicer's design and LCM operations, making it ideal for scenarios such as temporary or pop-up networks. In the H2020 5G-VICTORI project, this innovative approach was employed to facilitate media services in transportation verticals using a nomadic 5G node.
Future work includes the support of several virtualization instances per slice to allow UPF selection strategies such as load-balancing or edge computing, and the dynamic management of radio and compute resources according to slice requirements and traffic dynamics, including disaggregated RAN scenarios and the integration with O-RAN. 

\section{Acknowledgment}
This work was funded by the European Commission through the H2020 project 5G-VICTORI (grant agreement No. 857201). This work was also supported by the Spanish Ministry of Economic Affairs and Digital Transformation and the European Union – NextGenerationEU, in the framework of the Recovery Plan, Transformation and Resilience (PRTR) (Call UNICO I+D 5G 2021, ref. numbers TSI-063000-2021-12 – 6GENABLERS-DLT and TSI-063000-2021-15 – 6GSMART-EZ). The authors acknowledge CERCA Programme/Generalitat de Catalunya for sponsoring part of this work.

\bibliographystyle{IEEEtran}
\bibliography{references}

@IEEEtranBSTCTL{IEEEexample:BSTcontrol,
CTLuse_forced_etal       = "yes",
CTLmax_names_forced_etal = "2",
CTLnames_show_etal       = "1" }

@article{PAPAGEORGIOU2020232,
title = {{On 5G network slice modelling: Service-, resource-, or deployment-driven?}},
journal = {Computer Communications},
volume = {149},
pages = {232-240},
year = {2020},
issn = {0140-3664},
doi = {https://doi.org/10.1016/j.comcom.2019.10.024},
author = {Apostolos Papageorgiou and Adriana Fernández-Fernández and Shuaib Siddiqui and Gino Carrozzo}
}

@INPROCEEDINGS{9851288,
  author={Koné, Benjamin and Kora, Ahmed Dooguy and Niang, Boudal},
  booktitle={2022 45th International Conference on Telecommunications and Signal Processing (TSP)},
  title={{Network Resource Management and Core Network Slice Implementation: A Testbed for Rural Connectivity}},
  year={2022},
  volume={},
  number={},
  pages={200-205},
  doi={10.1109/TSP55681.2022.9851288}}

@Article{s21238103,
AUTHOR = {Fernández-Fernández, Adriana and Colman-Meixner, Carlos and Ochoa-Aday, Leonardo and Betzler, August and Khalili, Hamzeh and Siddiqui, Muhammad Shuaib and Carrozzo, Gino and Figuerola, Sergi and Nejabati, Reza and Simeonidou, Dimitra},
TITLE = {{Validating a 5G-Enabled Neutral Host Framework in City-Wide Deployments}},
JOURNAL = {Sensors},
VOLUME = {21},
YEAR = {2021},
NUMBER = {23},
ARTICLE-NUMBER = {8103},
PubMedID = {34884106},
ISSN = {1424-8220},
DOI = {10.3390/s21238103}
}

@INPROCEEDINGS{9912617,
  author={Chiu, Yi-Sung and Yen, Li-Hsing and Wang, Tse-Han and Tseng, Chien-Chao},
  booktitle={2022 IEEE International Conference on Service-Oriented System Engineering (SOSE)},
  title={{A Cloud Native Management and Orchestration Framework for 5G End-to-End Network Slicing}},
  year={2022},
  volume={},
  number={},
  pages={69-76},
  doi={10.1109/SOSE55356.2022.00014}}

@techreport{3gpp.28.801,
  author = {3GPP},
  month = Jan.,
  number = {28.801},
  title = {{Telecommunication management; Study on Management and Orchestration of Network Slicing for Next Generation Network}},
  type = {Technical Report (TR)},
  year = {2018}
}

@techreport{3gpp.28.533,
  author = {3GPP},
  month = Mar.,
  number = {28.533},
  title = {{Management and orchestration; Architecture framework}},
  type = {Technical Specification (TS)},
  year = {2023}
}

@ARTICLE{surveyslicing,
  author={Foukas, Xenofon and Patounas, Georgios and Elmokashfi, Ahmed and Marina, Mahesh K.},
  journal={IEEE Communications Magazine}, 
  title={{Network Slicing in 5G: Survey and Challenges}},
  year={2017},
  volume={55},
  number={5},
  pages={94-100},
  doi={10.1109/MCOM.2017.1600951}}

@techreport{ORAN-slicing,
  author = {{O-RAN Work Group 1}},
  institution = {O-RAN Alliance},
  month = {June},
  number = {10.0},
  title = {{O-RAN Slicing Architecture}},
  type = {Technical Specification (TS)},
  year = {2023}
}

@techreport{3gpp.23.501,
  author = {3GPP},
  institution = {{3rd Generation Partnership Project (3GPP)}},
  number = {23.501},
  title = {{System architecture for the 5G System (5GS)}},
  type = {Technical Specification (TS)},
  year = {2023}
}

@techreport{5gvictori,
  author = {5G-VICTORI},
  month = {March},
  title = {{D4.2: Intra-Field trials integration and vertical services execution and KPI validation }},
  type = {Project Deliverable},
  year = {2023}
}

@INPROCEEDINGS{amarisoft,
  author={Arroyo-Giganto, J. and González-Méndez, P. and Purriños-Paz, M. and Escudero-Garzas, J.J. and Losada-Sanisidro, P. and Fernández, Zaloa and Mogollón, Felipe and Yeregui, Inhar and Catalan-Cid, Miguel},
  booktitle={2022 3rd International Conference on Communication, Computing and Industry 4.0 (C2I4)}, 
  title={{Prototyping gNBs for Non-Public Networks}},
  year={2022},
  volume={},
  number={},
  pages={1-6},
  doi={10.1109/C2I456876.2022.10051296}}

\end{document}